# Graphene transistors for bioelectronics

Lucas H. Hess, Max Seifert, and Jose A. Garrido

*Abstract*—This paper provides an overview on graphene solution-gated field effect transistors (SGFETs) and their applications in bioelectronics. The fabrication and characterization of arrays of graphene SGFETs is presented and discussed with respect to competing technologies. To obtain a better understanding of the working principle of solution-gated transistors, the graphene-electrolyte interface is discussed in detail. The in-vitro biocompatibility of graphene is assessed by primary neuron cultures. Finally, bioelectronic experiments with electrogenic cells are presented, confirming the suitability of graphene to record the electrical activity of cells.

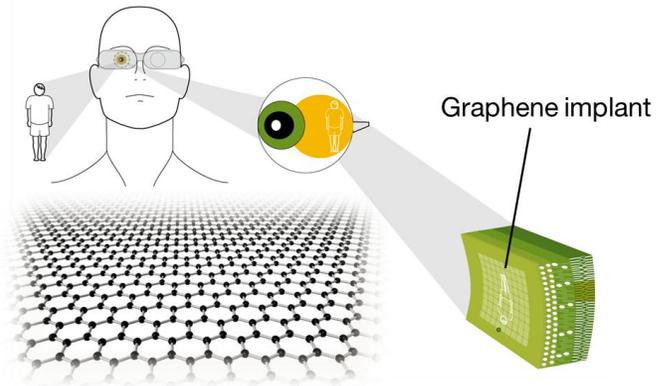

Fig. 1: Concept for a retinal implant. An image is acquired by a camera which is mounted to eyeglasses. After processing, the information is transferred to a retinal implant which stimulates nerve cells to transmit the signal to the brain.

## I. INTRODUCTION

A MAJOR challenge in the field of bioelectronics is the advancement of neural prostheses that allow restoring damaged abilities such as hearing[1] and vision, [2] or can help to find solutions for treating motor disabilities[3] or brain pathologies.[4] For this, it is essential to develop suitable interfaces between the biological system, such as the cochlea in the ear or the retina in the eye, and electronic devices. While the intermediate signal processing can be done easily with standard microprocessors, an efficient signal transfer from the electronics to the nervous systems, and vice versa, remains challenging.

Commercially available technologies are mostly based on micro electrode arrays (MEAs) made from silicon or metals. [1, 2, 5] MEA-based devices already have shown partial success in reconstituting hearing and vision, or in treating neural disorders. However, the performance of such implants, including long-term stability, is far from being satisfactory and has to be largely improved. In addition, MEA devices have an intrinsic poor spatial resolution, which is related to the rather large electrode impedance.[6]

Arrays of field effect transistors (FETs), on the other hand, have also been demonstrated for recording the electrical activity of nerve cells and tissue, [7, 8] and offer certain advantages with respect to MEAs such as the intrinsic amplification capabilities of FETs. Further, signal-to-noise ratios can be considerably enhanced if submicron FETs are used for recording. Finally, since FETs can be fabricated by standard semiconductor technology the production of high density structures is relatively easy, which has been used to demonstrate the

recording of cell activity with unprecedented spatial resolution.[9]

However, the presently used FETs platforms suffer from some major drawbacks, which are not only related to an imperfect technology but are also inherent to the materials that are employed.

One of the problems is the poor stability of many materials in the harsh biological environment. For example, devices fabricated from the ubiquitous semiconductor silicon are subject to stability problems in aqueous environments.[10, 11] This does not only result in a decreased performance of the sensor but may also damage the surrounding tissue.

Furthermore, to obtain high performance electronic devices exhibiting high charge carrier mobilities and low electronic noise, which are required for a high signal-to-noise ratio, it is necessary for classic semiconductors to build devices on the basis of highly crystalline substrates or even single-crystals. However, such good crystalline properties come hand in hand with mechanical characteristics like high rigidity and sharp edges. In biological systems, it is known that rigid and sharp prostheses can induce scarring around the device making it inoperative and damaging the surrounding tissue.

Thus, it is a major challenge to develop sensors from a chemically stable material which should exhibit good electronic properties allowing at the same time the fabrication of small, flexible devices. Graphene complies with all of these requirements.[12] Built up only from sp²-hybridized carbon

Manuscript received July 6, 2012. This work was supported by the European Union through the project NeuroCare, the German Research Foundation (DFG) in the framework of the Priority Program 1459 "Graphene", and the Nanosystems Initiative Munich (NIM).

The authors are with the Walter Schottky Institut, Technische Universität München, Am Coulombwall 4, 85748 Garching, Germany (e-mail: hess@wsi.tum.de; seifert@wsi.tum.de; garrido@wsi.tum.de).

TABLE I
OVERVIEW OF MATERIALS FOR SGFETS

| Material | Charge carrier mobility $\mu$ (cm²/Vs) | Interfacial capacitance $C_{int}$ ($\mu$F/cm²) | Transconductance $g_m/U_{DS}$ mS/V | Biocompatibility | Flexibility |
|---|---|---|---|---|---|
| Silicon[18] | 450 | 0.35 | 0.20 | 0[8] | 0[21] |
| Diamond[18] | 120 | 2 | 0.29 | +[22] | - |
| AlGaN/GaN[18] | 1200 | 0.32 | 0.51 | +[23] | - |
| Graphene[18] | 4000 | 2 | 5 | +[24] | +[20] |

**Table I: Properties of field effect transistors prepared using different material systems which have been proposed for bioelectronic applications. Graphene has both high charge carrier mobilities and a high interfacial capacitance which results in a high transconductive sensitivity. Additionally, graphene is chemically very stable, biocompatible, and is fully compatible with flexible technology**

atoms, whose interatomic bond is one of the strongest in nature, graphene shows a very high chemical stability even in harsh biological environments. As will be discussed later, this stability results in an excellent biocompatibility.

In addition, the band structure of graphene results in outstanding electronic properties.[13] Charge carriers close to the charge neutrality point, also referred to as Dirac point, behave like quasi-relativistic particles, which largely reduces scattering and gives rise to extremely high charge carrier mobilities. Even at near-room temperature, values of more than $10^5$ cm²V$^{-1}$s$^{-1}$ [14] can be reached with exfoliated graphene, outnumbering other semiconductors which have been so-far used for similar applications. Furthermore, the chemical properties of graphene allow the fabrication of transistors without any solid dielectric.[15, 16] As a result, interfacial capacitances of several $\mu$Fcm$^{-2}$ can be obtained, which is almost one order of magnitude higher than for traditional semiconductors with a fairly stable dielectric.[17] Consequently, graphene shows significantly higher transconductive sensitivities than other materials.[18]

Moreover, the fact that graphene is only one atomic layer thick and can be transferred to almost any arbitrary substrate – including thin polymer films – allows the fabrication of high performance fully flexible transistors.[19, 20]

Table I shows a comparison of the relevant properties of FETs based on graphene and other commonly used semiconductor materials. Although graphene technology is still in its infancy, graphene-based FETs already offer advantages when compared to the other competing technologies. In terms of biocompatibility, graphene exhibit a similar performance than diamond[22] and AlGaN/GaN,[23] whereas silicon devices require additional encapsulation layers to improve their stability.[25] Some materials, such as carbon nanotubes, also allow the fabrication of sensitive solution-gated transistors,[26] but their biocompatibility is still controversial.[27] Manufacturing flexible devices is not possible with single-crystalline diamond or AlGaN/GaN heterostructures and strongly deteriorates the electronic properties for silicon.[21] In contrast, graphene proved to be chemically stable and the feasibility of high-quality flexible transistors has already been demonstrated. [19,20]

Figure 2 shows a schematic of a neuroelectronic hybrid circuit based on graphene devices. Cells are grown on top of an array of graphene transistors, which can be used for the bidirectional communication with cells. In fact, the rather large interfacial capacitance of graphene SGFETs offers a more effective capacitive stimulation than Si-based SGFETs can provide. Thus, the graphene transistors can be used to either stimulate or record signals from cells. Recordings of the electrical activity of electrogenic cells using graphene transistors have been reported recently,[28,29] based on a configuration similar to the one shown in Figure 2, confirming the enormous potential of graphene SGFETs for bioelectronic applications.

In this paper, we summarize our work on graphene solution-gated transistors. Firstly, the basic working principle of graphene SGFETs is explained, including the fabrication and basic characterization. Then, the graphene/electrolyte interface is described in more detail to better understand the electrolyte-gating of graphene. Thereafter, the electronic requirements for semiconductor biosensors are discussed and the suitability of graphene SGFETs for these applications is assessed. In the final chapter, graphene SGFETs are tested in biological systems assessing their biocompatibility and demonstrating recordings of cell action potentials.

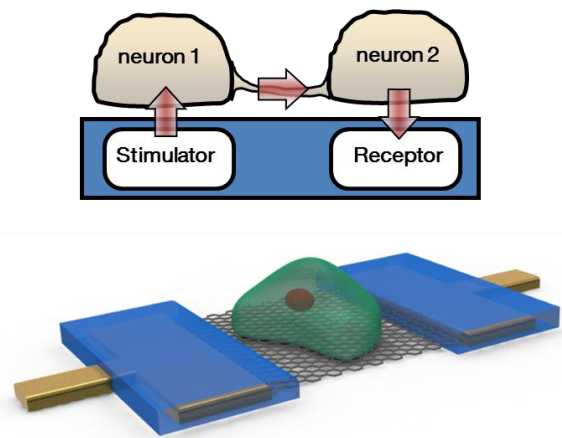

**Fig. 2: Neuroelectronic circuits and interfaces. For advanced neuroprostheses as well as for in-vitro neuronal networks studies, bidirectional communication with neurons is necessary, i.e. signals need to be transferred to and from the neurons to the electronics (upper part). Here, cells are grown on graphene transistors, which allow recording action potentials from the cells (lower part).**

## II. FABRICATION AND TECHNOLOGY

### A. CVD growth and transistor fabrication

High-quality graphene can be fabricated by several methods such as mechanical exfoliation, thermal decomposition of silicon carbide and chemical vapor deposition (CVD). Graphene sheets produced by mechanical exfoliation are very restricted in size (<1 mm²). Since in addition, the exfoliation procedure is typically done by hand,[30] this fabrication method is not suitable for industrial production.

Thermal decomposition of silicon carbide yields relatively large graphene sheets on insulating substrates[31] but it has major drawbacks related to this substrate. Firstly, SiC is rigid and hence not suitable for brain implant applications, where flexibility is an important issue. For this reason, it would be necessary to transfer the graphene to flexible substrates, which implicates the development of such a transfer process. Secondly, as the graphene is grown directly from the silicon carbide, a very strong electronic interaction between this substrate and the graphene is observed, which typically results in low carrier mobilities,[18, 31] unless special post-processing hydrogen treatments are applied.[32]

Graphene grown by CVD, on the other hand, appears as a more suitable alternative. It can be produced on a large scale[33, 34] at a relatively low cost, yielding high quality films with mobilities >40000 cm²V⁻¹s⁻¹.[35] Furthermore, it can be transferred to any substrate including flexible materials.

To grow graphene by CVD, a copper foil is introduced into a furnace(see Figure 3a) and is heated to 1000°C under hydrogen flow to remove the native copper oxide. After this etching step, the gas flow is changed to a methane/hydrogen mixture, which provides the carbon source for the growth of graphene on copper.

Subsequently, the graphene has to be removed from the copper and transferred to an insulating substrate. As depicted in Figure 3b, a layer of poly(methyl 2-methylpropenoate) (PMMA) is spin-coated on the graphene/copper stack to provide mechanical stability to the graphene layer. This stack is placed on the surface of an iron(III)chloride solution, which etches the copper under the graphene. After diluting, the graphene/PMMA layer is fished on the final substrate and the PMMA is removed by dissolving it in solvents. A final annealing step can be used to further remove residues from the PMMA.

Then, the graphene is patterned by optical lithography and an oxygen plasma etch is used in order to define the active area of the devices (Figure 3c). The ohmic drain and source contacts are prepared by gold evaporation and etching in a KI/I₂ solution. Finally, a structured layer of chemically stable SU8 photoresist is applied to the sample to protect the gold contacts from the electrolyte and to prevent leakage currents.

After a final annealing step, the devices are wire-bonded to a chip-carrier. The bonding wires are covered with silicone glue to insulate them from the electrolyte.

In order to have spatial resolution for the cell experiments, several transistors are fabricated in one array. The optical micrograph in Figure 3d shows one half of a 4×4 transistor array. For cell experiments, it is important that the size of the transistors, i.e. its gate area, is similar to the cell size. Large transistors, which are only partially covered by the cell, result in a poor gating of the device. In our experiments, typical transistors have a length of 10 µm and a width of 20 µm.

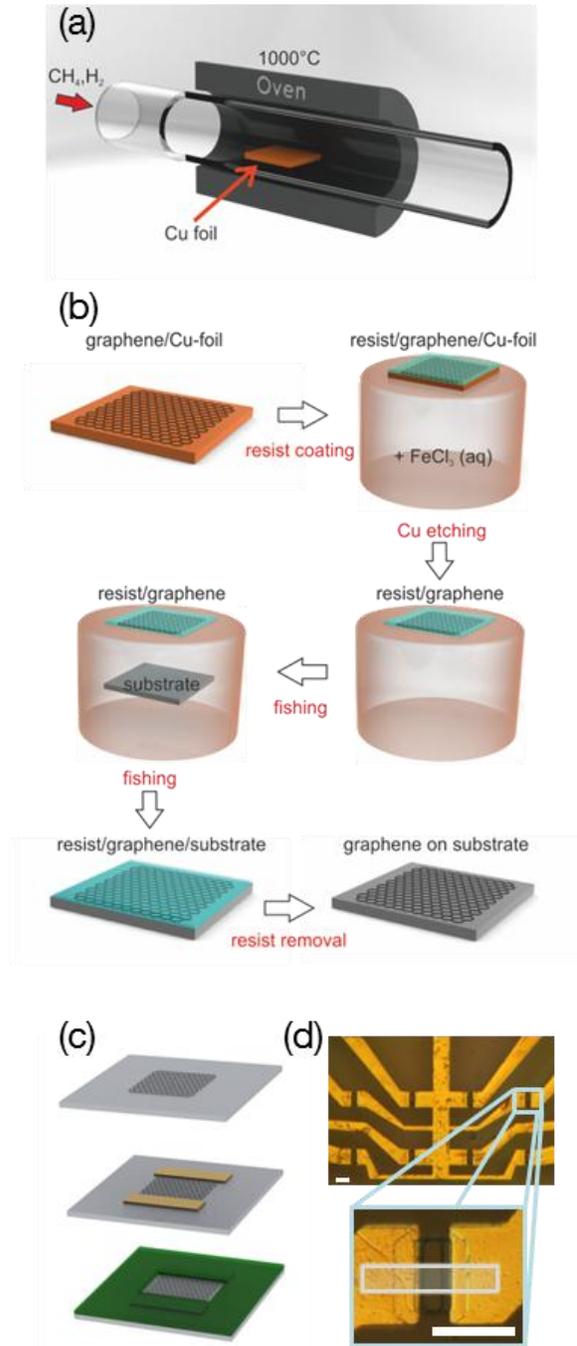

**Fig. 3: Fabrication of graphene SGFETs. a) Graphene films are grown on copper foil by CVD in a furnace at 1000°C under a flow of methane and hydrogen. b) With the help of a protective PMMA layer, the copper is etched away and the graphene is transferred to an insulating substrate. c) To fabricate the transistors, the graphene is etched with oxygen plasma and contacted via gold leads. For operation in an electrolyte, the metal is insulated with a chemically resistive resin. d) Optical micrograph of a transistor array and close-up of a single transistor. The graphene cannot be seen directly and is indicated with a box. The scale bars are 50 μm.**

### B. Basic characterization

For the characterization in electrolyte, the transistor array is immersed in a solution containing 5 mM phosphate buffered saline (PBS) and adjusted to an ionic strength of 100 mM with KCl. The gate voltage $U_{GS}$ is applied between the source contact of the transistors and a Ag/AgCl reference electrode in the solution. When a voltage is applied between the drain and the source contacts of a transistor, a current can be observed. This current is proportional to the applied drain-source voltage $U_{DS}$ (see Figure 4a) and can be modulated with the gate voltage. At a certain $U_{GS}$, the current shows a minimum value and increases almost linearly afar from this point (see Figure 4b).

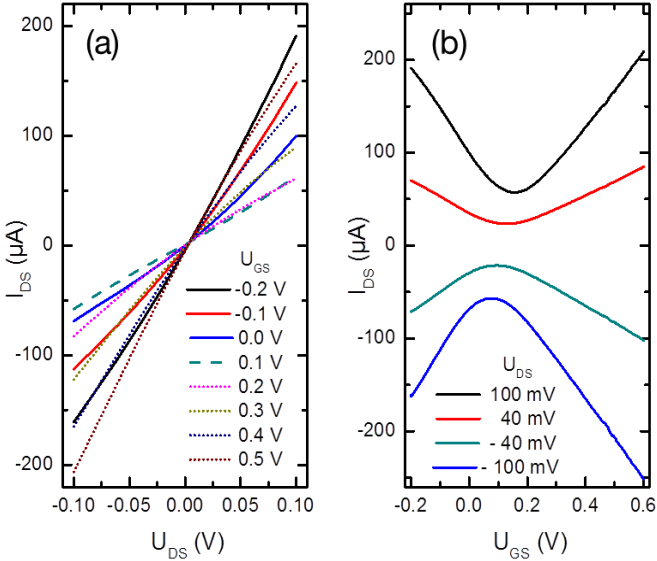

**Fig. 4: Current vs. voltage characteristics of a graphene SGFET. a) The drain-source current shows a linear dependence on the drain-source voltage. No saturation can be observed within the applied $U_{DS}$ voltages. b) When plotted against the gate-source voltage, the current shows a minimum and increases continuously afar from this point. The shown transistor is 20 µm wide and 10 µm long.**

## III. GRAPHENE-ELECTROLYTE INTERFACE

### A. Electrolyte Gating

The conductivity gating observed in Figure 4 can be explained by a simple model describing the graphene-electrolyte interface, as shown in Figure 5.[36] As the potential level of the Ag/AgCl electrode is fixed with respect to the vacuum level, applying a voltage between this electrode and the graphene shifts the position of the Fermi level in the graphene, which controls the number of free carriers induced electrostatically. When the Fermi level reaches the Dirac point, i.e. the energy where conduction and valence band meet, the conductivity in the graphene film shows its minimum value. The voltage at which this is observed is referred to as Dirac voltage $U_D$. For $E_F$ below $U_D$, the majority charge carriers are holes in the graphene valence band. When the gate voltage is further decreased, the Fermi level is shifted deeper into the valence band and the charge density increases. For $E_F$ above $U_D$, the conductivity is due to electrons in the conduction band, whose density can be modulated similarly. The voltage at which the Dirac point is reached in our experiments depends on several factors, such as the absolute electrochemical potential of the reference electrode and the work function of graphene. As both have similar values (4.6 eV [37, 38] and 4.5 eV [39], respectively), the expected Dirac voltage would be much closer to zero than the observed values between 0.3 V and 0.4 V. The reason for this difference is most likely due to the p-type doping of the graphene induced by its environment (such as the underlying substrate) or surface contamination (adsorbed water or remaining contamination from the device processing).[40] We have found a similar doping level for a variety of substrates including sapphire, silicon dioxide, and polyimide films, whereas n-type doping was observed for SGFETs prepared using graphene grown on SiC. [18]

Hall effect experiments employing solution-gated van-der-Pauw structures and Hall bars can be used to investigate the electrolytic gating of the carrier density in graphene.[36] Figure 6a shows the sheet conductivity as a function of the gate voltage obtained from these experiments revealing the expected V-shape curve typical of ambipolar transport, similar to the transistor curves shown in Fig. 4. Regarding the charge carrier density, a positive charge is observed for gate voltages below the Dirac point. As shown in Figure 6, the density of positive charge carriers decreases when $U_{GS}$ approaches the Dirac point. For gate voltages higher than $U_D$, the sign of the charge carriers is inverted and their density increases again. From the sheet conductivity and the charge carrier density, it is possible to extract the mobility of the carriers in the graphene sheet (see Figure 6b). Mobilities higher than 8000 $cm^2V^{-1}s^{-1}$ can be observed close to the Dirac point. Increasing the carrier density results in a decrease of the mobility, which reaches values of 1100 $cm^2V^{-1}s^{-1}$ and 800 $cm^2V^{-1}s^{-1}$ for holes and electrons, respectively, for a carrier density of $5\times10^{12}$ $cm^{-2}$. The observed dependence of the mobility with the carrier density is consistent with a scattering mechanism related to surface polar phonons, as has been previously reported for CVD graphene on different polar substrates.[41] The difference between electron and hole carrier mobility is probably due to impurity doping by the underlying substrate with different scattering cross sections for electrons and holes.[42]

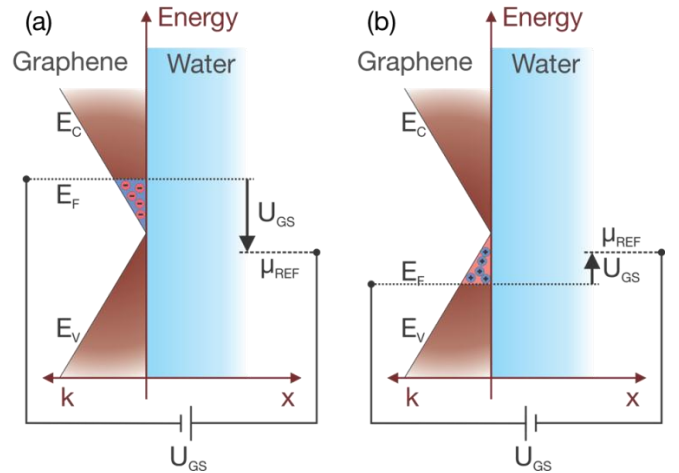

**Fig. 5: Electrolyte gating of graphene. By applying a voltage between the graphene and the reference electrode, the Fermi level in the graphene can be shifted. Hereby, the conductivity can be modulated and the type of charge carriers can be changed between electrons (a) and holes (b).**

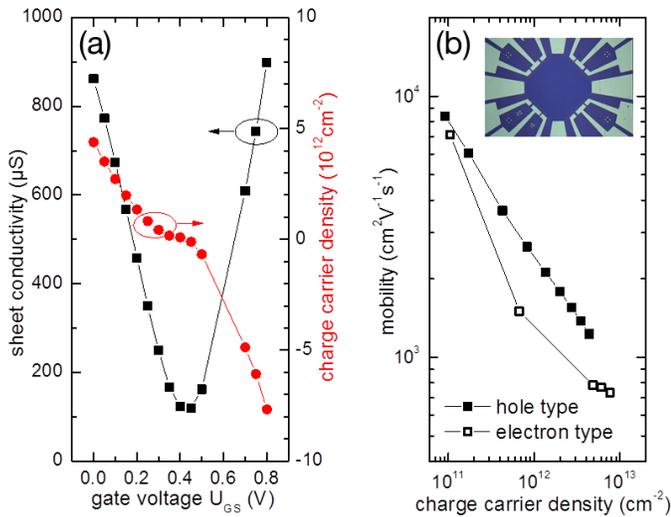

Fig. 6: **Electrolyte-gated Hall effect characterization. a) Sheet conductivity (black squares) and charge carrier density (red circles) are modulated with the gate voltage. At 0.4 V, the minimum conductivity, i.e. Dirac point, is reached and the sign of the majority carriers changes. b) Carrier mobility versus carrier density, revealing that holes exhibit higher mobilities than electrons. The mobility-density dependence suggests that carrier scattering is dominated by surface polar phonons.[41] Inset: Combined device with 8 transistors surrounding a van-der-Pauw structure.**

### B. Graphene-Electrolyte Interface

In order to obtain a quantitative description of the shift of the Fermi level and its effect on the conductivity modulation, the graphene-electrolyte interface must be considered in detail.

It is especially important to properly describe the interfacial capacitance which relates the charge carrier density in the graphene with the voltage applied between the graphene and the reference electrode, i.e. across the interface. This capacitance has two main contributions. Firstly, due to the density of states in graphene, the so-called quantum capacitance $C_Q$ has to be considered close to the Dirac point.[43] Secondly, the double layer capacitance formed at the graphene/electrolyte interface shall be considered.[36] On the one hand, the charge in the graphene side of the interface is easily described by the induced free holes and electrons, and is located in the graphene sheet. On the other hand, the type and position of charges in the electrolyte requires a more elaborated description. Several different types of cations and anions are typically present and can move freely in the electrolyte. Charges in the graphene and at its surface are compensated by electrolyte ions depending on the sign of the ions' charges and the graphene surface charge. Due to the consequent charge screening induced by electrolyte ions, the imbalance between differently charged ions will decrease deep into the electrolyte. In a first approximation, this decrease can be described by an exponential decrease afar from the interface.

However, close to the interface, the above approximation does not hold. Very close to hydrophobic surfaces such as graphene, the structure of the water itself changes drastically.[44] As shown by molecular dynamics simulations,[45] the density of water decreases strongly at the surface resulting in a so-called hydrophobic gap between the solid and the electrolyte (see Fig. 7). In this gap, the effective dielectric constant is much smaller than in bulk water resulting in a large potential drop at the interface.

The graphene/electrolyte interface was simulated with the commercially available Poisson-Schrödinger solver nextnano³.[46] This software self-consistently simulates the electronic structure of semiconductor materials as well as the potential and ion distribution in the electrolyte. On the electrolyte side, the above mentioned hydrophobic gap was considered with the help of a non-uniform dielectric constant of the water close to the surface which was obtained from molecular dynamics simulations. [45] The ion distributions were calculated with a Poisson-Boltzmann model which was extended by spatially varying potentials of mean force (PMF) for each ion; such PMFs are introduced to prevent the unphysical situation of ions approaching the surface infinitely close.[45]

Figure 8 shows the charge carrier density obtained experimentally as well as the results from the simulations. A good agreement is seen for the hole regime ($U_{GS}$-$U_D$<0). The experimentally observed hole-electron asymmetry is not included in the model so far.

The inset of Fig. 8 shows the interfacial capacitance obtained from the simulation using the extended Poisson-Boltzmann (ePB) model (blue curve). As discussed earlier, this capacitance is composed of two contributions, the quantum capacitance (grey curve) and the double layer capacitance of the interface, which includes both the potential drop due to the hydrophobic gap and the varying PMF for the ions. The simulations in Fig. 8 reveal that away from the Dirac pint, the interfacial capacitance is dominated by the contribution of the electrical double layer. In a first approximation, the double layer capacitance, about 2-3 µF/cm2, can be obtained when considering a hydrophobic gap of about 0.3 nm and with a dielectric constant of 1 (red line in inset of Fig. 8).

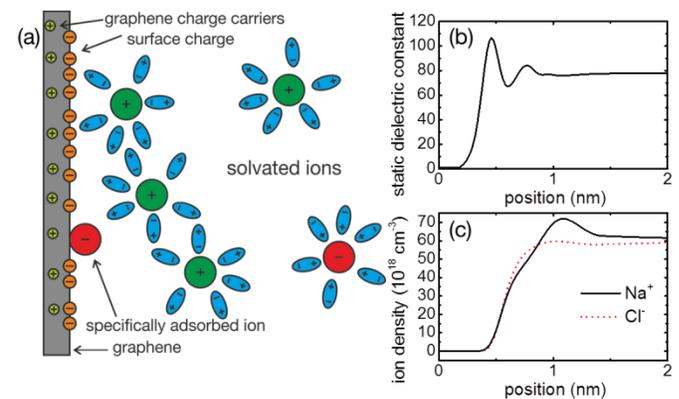

Fig. 7: **a) Schematic view of the graphene/electrolyte interface. Three different kinds of charges can be identified: Free carriers in the graphene, fixed surface charges and ions in the electrolyte. b) Effective dielectric constant of water close to the hydrophobic surface, as obtained from molecular dynamics simulations.[45] c) Simulated ion densities of sodium and chloride ions close to the interface. Due to negative interface charges, an abundance of positive ions is observed. (The simulation is obtained considering an interfacial charge of $10^{12}$ negative charges per cm², a salt concentration of 100 mM, a pH of 7, and zero gate biasing.)**

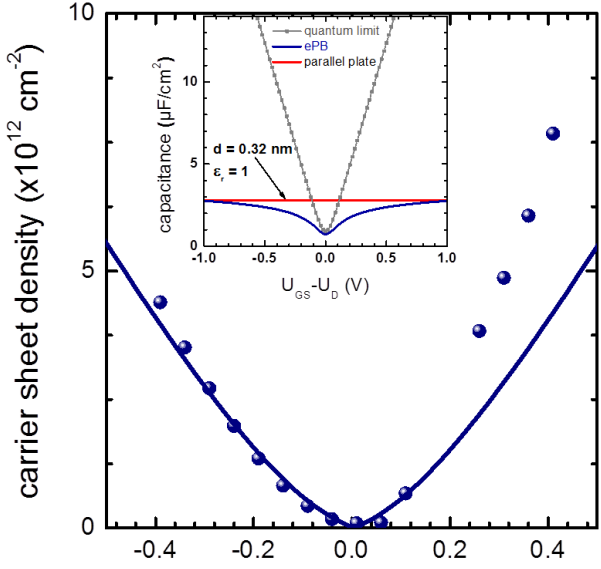

**Fig. 8: Comparison of simulated charge carrier densities (line) with experimental values. For holes, theoretical and experimental results are in good agreement, whereas a small difference can be seen for electrons. The inset shows a model for the interfacial capacitance. Both the quantum capacitance of the graphene and the semiconductor-electrolyte interfacial capacitance have to be considered.**

The description of the graphene/electrolyte interface presented above already anticipates the ion sensitivity of graphene SGFETs. The influence of the ion concentration and valence on graphene SGFETs is summarized in Figure 9, which shows that the addition of different salts with mono- and divalent cations (NaCl and CaCl$_2$) results in a shift of the Dirac point, i.e. a decrease in the Dirac voltage. This effect can be explained with the simple model[47] of Figure 9b, which shows the different potentials and charges relevant for describing the interface. Besides the charge in the graphene $\sigma^{gra}$ (resulting from the free carriers) and the ionic charge in the electrolyte $\sigma^{dif}$, a surface-bound charge can be assumed directly at the interface. To ensure charge neutrality, the sum of these charges must be zero. In this simplified picture, the important potentials are the potential of the graphene $\varphi^{gra}$, i.e. the Fermi level, the potential of the reference electrode $\varphi^{ref}$, which is fixed with respect to the vacuum, and the potential at the interface $\varphi^{dif}$. $\sigma^{gra}$ is defined by $\varphi^{ref}$ and the gate-source voltage. Furthermore, the charge in the graphene is related to the $\varphi^{dif}$ via the interfacial capacitance $C_H$. The charge and potential distribution shown in Fig. 9b correspond to a situation with a negatively applied gate voltage and a negative surface-bound charge.

The diffusive counter-charge $\sigma^{dif}$ in the electrolyte can be modeled in a first approximation using the Grahame equation:

$$\sigma^{dif} = \sqrt{2\epsilon\epsilon_0 RT} \left[ \sum_i c_i \cdot \exp\left( \frac{-z_i F \varphi^{dif}}{RT} \right) \right]^{1/2} \quad (1)$$

Where $c_i$ and $z_i$ are the ion concentration and valence for the different ions, $\epsilon$ and $\epsilon_0$ are the relative and the vacuum permittivity, $R$ is the general gas constant, $F$ is the Faraday constant, and $T$ is the absolute temperature.

If the ion concentration $c_i$ varies, the charges in the diffusive layer and the potential at the interface have to adapt so that Eq.1 is still satisfied. The charge in the graphene $\sigma^{gra}$ is directly connected to $\sigma^{dif}$ and $\varphi^{dif}$. Firstly, any variation in $\sigma^{gra}$ is related to $\sigma^{dif}$ through the charge neutrality condition; thus, $\partial\sigma^{gra} = -\partial\sigma^{dif}$. Secondly, $\sigma^{gra}$ is coupled to $\varphi^{dif}$ via the interfacial capacitance, such that

$$\partial\sigma^{gra} = -C_H \partial\varphi^{dif} \quad (2)$$

For the case depicted in Fig 9b, this means that an increase in the ion concentration results in a decreased charge in the graphene, which is consistent with the experimental results shown in Fig. 9a.

As can be seen from Eq. 1, the addition of divalent ions affects the interfacial potential in a different way than for monovalent ions, which can be explained by considering the effect of the ion valence.

Figure 9a shows the experimental comparison for mono- and divalent ions, which is in good agreement with the model. The flattening observed at low ionic strengths for both the experimental data and the model is due to the background ion concentration from the buffer.

To obtain the observed decrease in the Dirac voltage, a surface charge of -10$^{-5}$ C cm$^{-2}$ has to be assumed. Although the origin of this charge is not identified yet, it is probably due to charged groups at the substrate or related to defects in the graphene. Presumably, the pH sensitivity of graphene transistors, which has been observed by several groups,[48, 49] is also related to such surface groups that can be protonated/deprotonated, e.g. OH groups.[50]

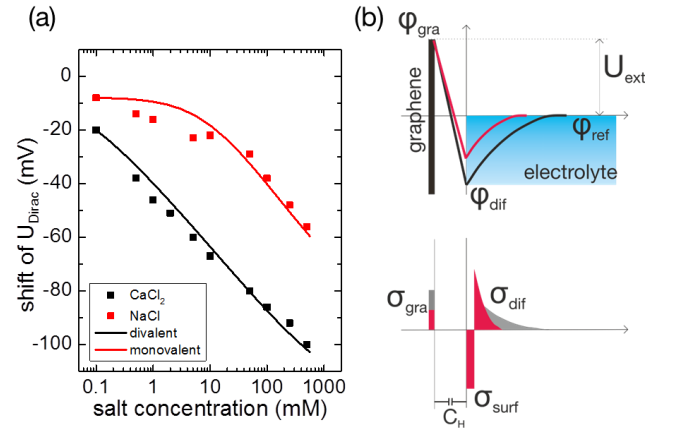

**Fig. 9: Ion sensitivity of graphene SGFETs. a) For increasing ion concentration, a shift of the Dirac point to more negative voltages is observed. At low concentrations, the measurements are influenced by the buffer's background concentration (in this case 5mM HEPES). The ion valence can explain the different response to monovalent and divalent ions (see Eq. 1). b) Potential profile (top) and charge distribution (bottom) at the graphene/electrolyte interface assuming U$_G$<U$_D$ and a negative surface charge (σ$_{surf}$) For higher ion concentration (colored in red), the diffusive charge is increased, resulting in a lower charge in the graphene layer, i.e. a shift of the Dirac point towards more negative voltages.**

## IV. TRANSISTOR PERFORMANCE

To compare the performance of graphene SGFETs to devices fabricated using other materials, it is important to understand the way typical cell experiments are performed. Figure 10a shows a typical configuration for the recording of cell action potentials. Constant biasing voltages ($U_{DS}$ and $U_{GS}$) are applied to the transistor and the drain-source current ($I_{DS}$) is monitored. If a small, local change of the potential at the surface occurs, e.g. induced by a cell action potential, a small variation in $I_{DS}$ can be observed. The ratio of this change in current to the original voltage signal at the surface is related to a fundamental property of the transistor: the transconductance. The transconductance $g_m$ is defined as the derivative of the drain-source current with respect to the gate-source voltage. Thus, in a device with a higher transconductance, a gate variation results in a higher current modulation compared to a low-transconductance device. Experimentally, higher currents are easier to detect and less prone to be disturbed by external noise, i.e. devices with a high transconductance can detect smaller signals and are therefore more sensitive.

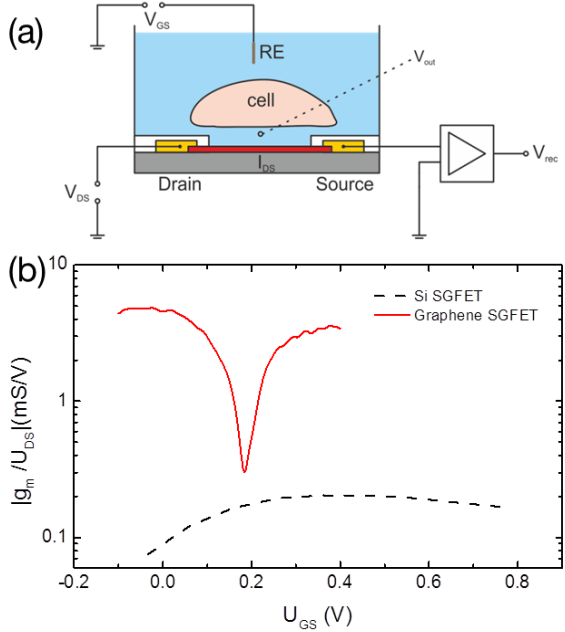

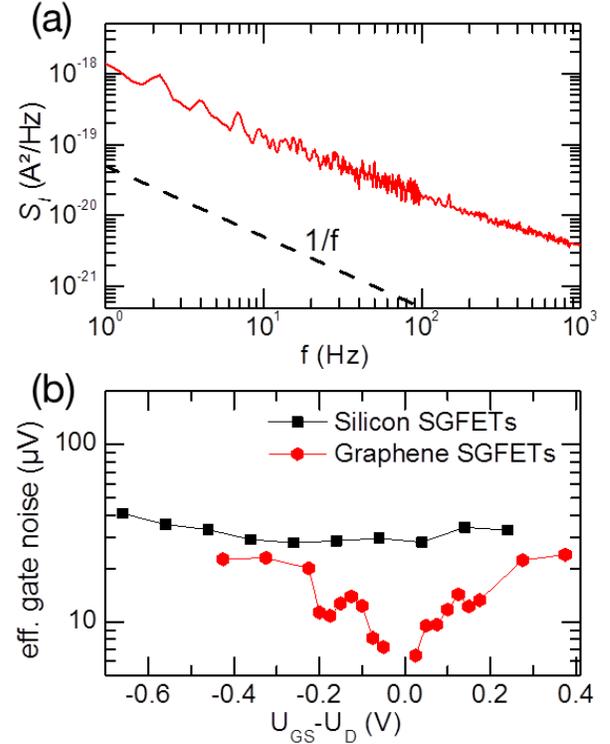

**Fig. 11:** a) Current noise power spectral density of a graphene SGFET revealing a *1/f* dependence. b) Effective RMS gate noise obtained from Equation 3. For graphene, values below 10μV can be reached, which is comparable to microelectrode arrays.

**Fig. 10:** a) Schematic view of a cell-transistor measurement. A local, small potential change triggered by the cell results in a modulation of the transistor current. The current is converted to a voltage and amplified. b) Comparison of the transconductance of graphene and Si SGFETs. The higher $g_m$ of graphene FETS results from both the higher charge carrier mobility and the higher interfacial capacitance (see Table I).

In a standard MOSFET model, the transconductance is proportional to the charge carrier mobility and the interfacial capacitance. Whereas the carrier mobility is a material parameter, the interfacial capacitance depends on the device design. In the case of Si SGFETs based on a MOSFET concept, the interfacial capacitance depends on the thickness and material of the insulating gate dielectric. For graphene SGFETs, where no such dielectric layer is used, the interfacial capacitance is governed by the electrical double layer at the graphene/electrolyte interface, as shown in Figure 8. Figure 10b compares the transconductance of a graphene SGFET with a MOSFET-based silicon device.[18] A clear dip can be seen in the transconductance of the graphene transistor which is consistent with the decrease of the interfacial capacitance at the Dirac point. It can be seen clearly that the graphene device outperforms the silicon transistor by more than an order of magnitude, which can be well understood when comparing the values of mobility and interfacial capacitance given in Table I. It is expected that with further development in graphene growth and technology higher mobilities and thereby higher transconductances can be reached.

The sensitivity of an SGFET is affected by random current fluctuations intrinsic to any metal or semiconductor, known as internal noise.[51] Figure 11 shows an exemplary power spectral density of the current noise of a graphene SGFET. As for many semiconductors and metals, an inverse dependence on the frequency can be observed for graphene SGFETs (*1/f* noise). For graphene, the origin of this noise is not clear yet. Possible sources of noise include charge trapping at the interface to the substrate as well as the so-called charge noise[52]. For silicon devices, the noise is induced by charge traps in the oxide. Carriers from the silicon can tunnel into these traps and lead to a certain charge density in the oxide. The scattering induced by this charge results in a *1/f* noise spectrum[53]. For graphene transistors, no such oxide is used, which is one reason for the observed lower noise.[52, 54]

To relate this current noise to the experimentally relevant voltage signals, the current noise can be converted to an effective gate voltage noise by dividing by the square of the transconductance:

$$(\Delta U_{GS}^{RMS})^2 = \int_{1Hz}^{10kHz} \frac{S_I}{g_m^2} df \quad (3)$$

Integrating over the typical frequency range for cell experiments (1 Hz- 10 kHz) yields an RMS value of less than 10 μV, which is comparable to state-of-the-art MEA devices used for the detection of action potentials from cells. This noise level is below the expected signal generated by the action potentials of mammalian neurons.[7, 55]

## V. Interfacing of Cells with Graphene

### A. Biocompatibility

For the development of sensors to be used in a biological environment, it is of crucial importance that neither the sensor nor the investigated biological systems is harmed by the respective component. As mentioned before, graphene exhibits a remarkable chemical stability in rather harsh aqueous environments. To investigate the response of sensitive living cells to the presence of graphene in their environment, primary retinal ganglion cells from rats (postnatal day 7 as well as 8 weeks old) were purified and seeded on glass slides and graphene-coated sapphire.[24] After four or six days (for the young and adult rats, respectively), the survival of these retinal neurons was investigated using fluorescence microscopy. Viable cells were labeled with calcein, a green fluorescent fluorophore. It was observed (see Figure 12) that both young and adult neurons survive and grow neurites on graphene and the standard glass substrate, confirming the excellent biocompatibility of the CVD graphene.[24]

To assess the influence of different substrates and of material boundaries on neuron viability, cells have been cultured on alternating graphene and sapphire structures. As can be seen from the scanning electron micrograph in Fig 12b, the cell growth was not affected by the graphene/sapphire patterned structure.

### B. Cardiomyocyte-like cells (HL-1)

Furthermore, cardiomyocyte-like HL-1 cells were cultured on the transistor arrays.[28] After several days in vitro, the cells formed a densely packed layer on the array and showed healthy growth. Figure 13a shows the combination of a differential interference contrast microscopy image and a fluorescence image. The fluorescence image reveals the cell layer with the help of the stain calcein which is metabolized by the healthy cells.

The bias of all transistors in the array was chosen to maximize the transconductance. At this fixed bias, the currents through the transistors were recorded over time. At certain times, the recordings show spikes in the current with a very small temporal difference for all transistors, as is depicted in Figure 13b and c. The current spikes can be converted to gate voltage spikes using the transistor transconductance; values up to 1 mV with an RMS noise level of 50 μV were recorded. These spikes can be attributed to the propagation of action potentials across the cell layer as expected for cardiomyocytes.[56] The temporal difference between the spikes on different transistors results from the propagation speed of action potentials across the cell layer. The shape of the signals varies for different transistors (see Figure 13d) depending on the coupling of the cells to the individual transistors.[55]

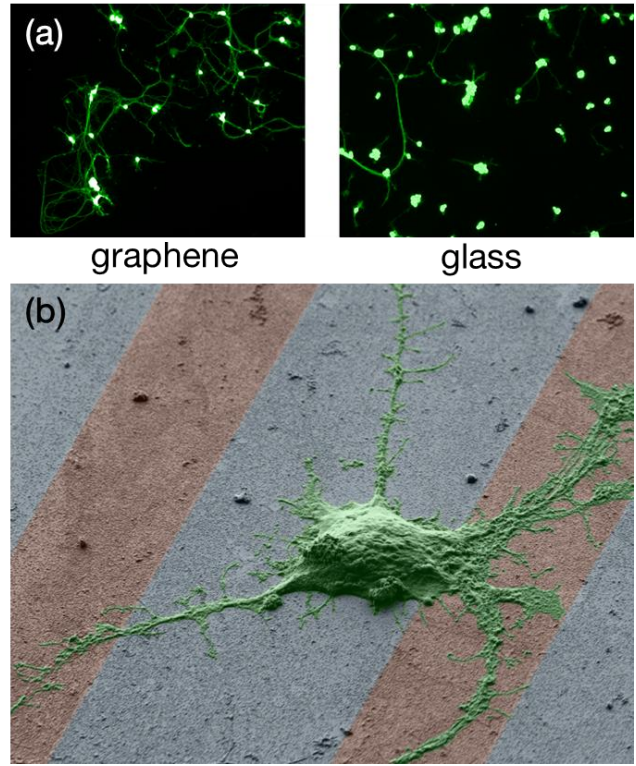

Fig. 12: Viability tests with retinal ganglion cells assessing the biocompatibility of CVD graphene. a) Retinal neurons are able to survive and grow neurites on both graphene and the reference substrate glass. (Courtesy of A. Bendali, Institut de la Vision, Paris, France) b) Scanning electron micrograph of a neuron on a patterned graphene/sapphire structure. Neither cells nor neurites show differences on an alternating graphene (10 μm wide, colored in red) and sapphire (20 μm wide, colored in blue) structure.[24]

### C. Human Kidney Cells (HEK-293)

In order to better understand the coupling between cells and the transistors, human embryonic kidney cells from the HEK-293 line have been grown on the transistor arrays. These cells have several properties that facilitate an investigation of the coupling. Particularly, HEK-293 cells can be used for patch-clamping experiments. In these experiments, a patch pipette containing a second electrode is introduced manually in the inner part of a single cell providing the possibility to apply voltages between the inside and the outside of the cell. Like this, specific voltage-gated ion-channels in the cell membrane can be opened or closed and the current across these channels can suggested that the ion concentration in the cleft between the cell and the transistor changes due to the additional ions released through the ion channels. As the rate at which ions can leave the cleft is limited by diffusion, it can take some time until the conflux and efflux of ions reach equilibrium.[60] This local and temporary change of the ion concentration influences the nearby transistor, as discussed in section III.

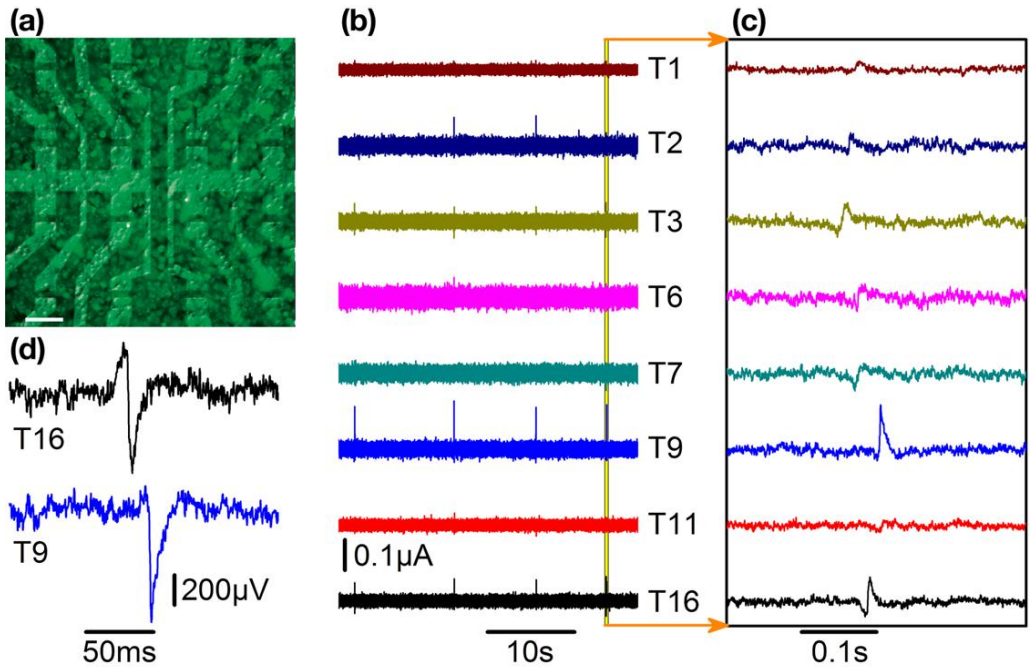

**Fig. 13: Electrical activity of HL-1 cells recorded with an array of graphene SGFETs. a) A combination of an optical micrograph of an array of 16 transistors and a fluorescence image of the stained cells on the array. The scale bar is 100µm. b,c) Current recordings of 8 transistors of one array over different time ranges. d) Close-up of exemplary single spikes. The current signal has been converted to an equivalent gate-voltage signal. (Copyright 2011 Wiley. Used with permission from [28])**

## VI. CONCLUSION

In this article, we have provided some insight on fundamental aspects of graphene solution-gated field effect transistors, and at the same time we have discussed their use as transducers for the recording of the electrical activity of living cells.

We have shown that due to its outstanding chemical, electrical, and mechanical properties, graphene is an ideal material for the fabrication of bioelectronic devices based on field effect transistors. In particular, the high mobility of carriers in graphene together with the singular double layer formed at the graphene/electrolyte interface results in FET devices which far outperform current technologies in terms of their gate sensitivity. Further, even at this relatively early stage of development, graphene FETs exhibit a noise performance that equals or even surpasses that of already well-established technologies. New advancements in the growth of high quality graphene are expected to further increase the substantial advantages of graphene for sensing applications.

Looking beyond the state of the art described in this article, the challenge resides in the development of high performance graphene-based devices on flexible substrates. Provided that this issue can be successfully addressed, graphene-based SGFETs have the potential to set a new paradigm in bioelectronics and, in particular, in the field of neural prostheses.

be recorded.[57] The cells used in our experiments are genetically modified in such a way, that their membrane contains only one type of ion channels, in our case potassium channels.[58] Therefore, all observed currents and transistor signals can be attributed to these channels which simplifies the analysis significantly.

Figure 14 shows the results of such a patch-clamp experiment. A certain voltage sequence is applied across the cell membrane resulting in the opening of the potassium channels. The voltage-gating of the $K^+$-channels is confirmed by measuring the current across the membrane with the help of the patch electrode (red curve in Fig. 14c) as well as by the response of the transistor (blue curve in Fig. 14c). The outflow of $K^+$ ions resulting from the cell stimulation induces an ionic current along the cleft formed between cell and transistor. Such a current translates into a local variation of the effective gate voltage of the underlying transistor. [8, 59] Figure 14c shows a single recording (light blue curve) from a transistor, which exhibits a signal-to-noise ratio of 13 with an RMS noise of 23 µV. As this kind of membrane currents is externally triggered with the patch pipette, the same process can be repeated several times yielding the same result. By averaging the transistor signals, it is possible to obtain a signal-to-noise ratio of 50. When the transistor signal is compared to the membrane current, small differences can be identified after the potential has changed. Both after opening and closing the ion channels, the membrane current changes faster than the transistor signal. This suggests that the coupling is not merely due to the charge transferred across the membrane. It has been

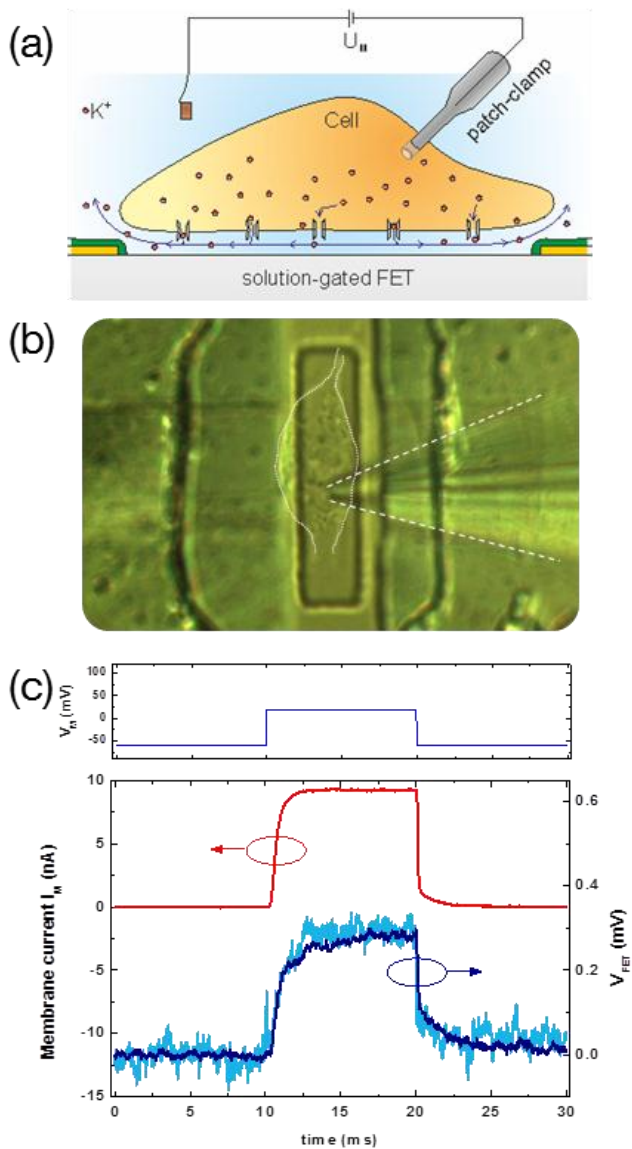

**Fig. 14: Single cell coupling to a graphene transistor. a) Schematic of a patch-clamp experiment representing the voltage-gated opening of K-ion channels via an external glass pipette and the ionic current flowing in the cleft between cell and transistor. b) Optical micrograph of a patched-clamped HEK293 cell on a graphene transistor. For better visibility, the cell and the pipette are marked with white lines. c) Voltage sequence applied to the patch electrode (top) and the resulting membrane current (middle) and transistor recording (bottom). The light blue line shows the original data whereas the dark blue shows an averaged signal.**

## Acknowledgment

We would like to thank our collaboration partners M. Jansen, V. Maybeck, and A. Offenhäusser from the Peter Grünberg Institute and Institute of Complex Systems at the Forschungszentrum Jülich, Germany, as well as A. Bendali and S. Picaud from the Institut de la Vision, Paris, France. Furthermore, we would like to acknowledge the work done by C. Becker-Freyseng, B. Blaschke, M. Hauf, S. Matich, M. Dankerl, S. Birner, I.D. Sharp, and M. Stutzmann at the Walter Schottky Institute.